\documentclass[8pt]{article}
\usepackage[utf8]{inputenc}
\usepackage{indentfirst} 

\usepackage[english]{babel}

\usepackage[a4paper,top=3cm,bottom=2cm,left=3cm,right=3cm,marginparwidth=1.75cm]{geometry}

\usepackage{amsmath,amssymb,amsthm}
\usepackage{algorithm}
\usepackage{algpseudocode}
\usepackage{graphicx}
\usepackage{bbm,enumerate}
\usepackage{xcolor}
\usepackage[authoryear,round]{natbib}
\usepackage{setspace}
\usepackage[colorlinks=true, allcolors=blue]{hyperref}
\usepackage{bbm}
\usepackage{comment}
\usepackage[export]{adjustbox}
\usepackage{subcaption}
\usepackage{authblk}

\theoremstyle{plain}

\title{Optimal Sample Splitting for Observational Studies}
\author[1]{Qishuo Yin}
\author[2]{Dylan S. Small}
\affil[1]{Department of Operations Research and Financial Engineering, School of Engineering and Applied Sciences, Princeton University}
\affil[2]{Department of Statistics and Data Science, The Wharton School, University of Pennsylvania}
\date{Jan 29 2026}

\begin{document}
\setstretch{2}

\maketitle

\begin{abstract}
In observational studies of treatment effects, estimates may be biased by unmeasured confounders, which can potentially affect the validity of the results. Understanding sensitivity to such biases helps assess how unmeasured confounding impacts credibility. The design of an observational study strongly influences its sensitivity to bias. Previous work has shown that the sensitivity to bias can be reduced by dividing a dataset into a planning sample and a larger analysis sample, where the planning sample guides design decisions. But the choice of what fraction of the data to put in the planning sample vs. the analysis sample was ad hoc.  Here, we develop an approach to find the optimal fraction using plasmode datasets.  We show that our method works well in high-dimensional outcome spaces.  We apply our method to study the effects of exposure to second-hand smoke in children. The \texttt{OptimalSampling} R package implementing our method is available at \href{https://github.com/qishuoyin/Optimal_Sampling}{GitHub}.
\end{abstract}


\section{Introduction}
\label{s:introduction}
Causal inference is easiest in a randomized experiment because each unit has a known probability of assignment to treatment or control, and this provides a basis for inference  \citep{fisher1937design}. However, randomized experiments are sometimes infeasible or unethical, and observational studies must be considered. In observational studies, the treatment probabilities are unknown, and there may be unmeasured confounders \citep{dorn1953philosophy}.  

Matching methods, inspired by stratified randomization in randomized experiments, enable randomization inference methods to be used if there are no unmeasured confounders. However, there may be unmeasured confounders which can cause treatment effect estimates to be biased. Sensitivity analysis can be used to measure how sensitive findings (that assume no unmeasured confounding) are to bias from unmeasured confounders.

The sensitivity of an observational study to bias can be substantially affected by its design \citep{rosenbaum2004design}. For example, if there are multiple outcomes, choosing certain of the outcomes to focus on can substantially decrease sensitivity to bias as discussed in \cite{heller2009split}. The paper introduced a sample splitting approach in which a pilot sample is used to choose the design, and then the remaining part of the sample is used for analysis. They showed that the approach can substantially reduce sensitivity to bias. However, its performance can vary significantly depending on the chosen sample split fraction. In this paper, we propose a method that determines the optimal sample splitting fraction which is adaptive to both low and high dimensional outcomes space. Our method numerically optimizes the sample splitting fraction for a semi-synthetic data set.  We also provide implementation guidance for the practitioner with empirically recommended splitting fractions that have close to optimal performance across a range of settings.

We apply our method to assess the health effects of children exposed to second-hand smoke. Such an assessment must rely on observational data due to ethical constraints preventing randomized trials \citep{shaper1982effects, grasmick1985combined, smith2003effect}. \cite{mannino2003second} analyzed the relationship between second-hand smoke exposure and blood lead levels using the Third National Health and Nutrition Examination Survey. More recent work \cite{braun2020influence} suggests that exposure to second-hand smoke influences multiple biological pathways beyond blood lead levels. Because this evidence is from observational studies and there may be unmeasured confounders, a sensitivity analysis is useful. 

Following previous studies \citep{mannino2003second}, \cite{zhang2020calibrated} of the health effects of children's exposure to second-hand smoke, we used matching to control for aspects of a child's demographics and socioeconomic status (gender, race/ethnicity, age, education of the reference adult in the child's household, poverty level of the child's family, number of rooms in the house and family size). However, there may still be unmeasured covariates that we may fail to take into consideration when conducting matching. Then, each child in a matched pair may not live in exactly the same environment as the other in the same pair. To address this, we performed a sensitivity analysis to examine how children’s health is influenced by second-hand smoke under different bias factors in the assignment of treatment. This sensitivity analysis will give a conclusion about which outcomes are affected by second-hand smoke even allowing for some unmeasured confounding. Section \ref{s:application} presents the detailed results. 

Sensitivity analysis plays a central role in assessing the robustness of causal conclusions in observational studies. Various sensitivity analysis methods have been developed for different contexts \citep{rosenbaum1983a, rosenbaum1983b, rosenbaum1987, rosenbaum2004, greenland2005, vanderweele2017, vanderweele2019, mathur2020, yin2022, zheng2023}. 
We develop our method for Rosenbaum's sensitivity analysis framework, which is particularly useful for matched observational studies. Our approach provides a unified framework that provide the solution of optimal sample splitting fractions, accommodates high-dimensional outcome spaces through multiple testing procedures, and improves design sensitivity in observational studies.                    

This paper proceeds as follows. Section \ref{s:notation} introduces the notation and sensitivity model. Section \ref{s:method} develops the optimal sample split method for multiple outcome tests. Section \ref{s:simulation} presents the simulation results in different settings. Finally, Section \ref{s:application} applies our methodology to study the health effects of second-hand smoke on children.

\section{Notation and Review}
\label{s:notation}
\subsection{Paired Randomized Experiments}
\label{s:preliminary_paired}
Consider $I$ matched pairs of two individuals, indexed by $j = 1, 2$ within each pair $i = 1, \dots, I$. We analyze $K$ outcomes of interest, indexed by $k = 1, \dots, K$. Let $\mathbf{x}_{ij}$ denote the observed pretreatment covariates, $u_{ij}$ the unobserved covariates, and $Z_{ij}$ the observed treatment assignment. Define the observed response vector for the $k^{\text{th}}$ outcome as $\mathbf{R}_{k} = \left( R_{11k}, R_{12k}, \dots, R_{I1k}, R_{I2k} \right)^T$.

Treatment assignment is binary: $Z_{ij} = 1$ indicates treatment, and $Z_{ij} = 0$ indicates control. Within each matched pair, exactly one individual receives treatment and the other receives control, ensuring $Z_{i1} + Z_{i2} = 1$ for all $i$. Matching is conducted on observed covariates, i.e., $\mathbf{x}_{i1} = \mathbf{x}_{i2}$, though unobserved confounders may differ, i.e., $u_{i1} \neq u_{i2}$.

For each individual in pair $i$, the potential outcomes are denoted as $r_{T_{ik}}$ (under treatment) and $r_{C_{ik}}$ (under control). The $k^{\text{th}}$ observed response is therefore:
\begin{align*}
    R_{ijk} = Z_{ij} r_{T_{ik}} + (1 - Z_{ij}) r_{C_{ik}}, \quad k = 1, \dots, K.
\end{align*}
Define $\mathbf{r}_{C_k} = (r_{C_{1k}}, \dots, r_{C_{Ik}})^T$ and $\mathbf{r}_{T_k} = (r_{T_{1k}}, \dots, r_{T_{Ik}})^T$. The treatment effect for pair $i$ on outcome $k$ is given by:
\begin{align*}
    \delta_{ik} = r_{T_{ik}} - r_{C_{ik}}.
\end{align*}
Fisher's sharp null hypothesis \citep{fisher1937design} for outcome k asserts that the treatment has no effect on outcome $k$ for all subjects:
\begin{align*}
    H_0^{k}: \delta_{ik} = 0, \quad \forall i \quad \text{ for outcome } k.
\end{align*}
Define the set of all possible treatment assignments as:
\begin{align*}
    \mathcal{Z} = \{ \mathbf{Z}  = ( Z_{11}, Z_{12}, \dots, Z_{I2}) \mid Z_{ij} \in \{0,1\}, \forall i, j \},
\end{align*}
which contains $|\mathcal{Z}| = 2^I$ elements. Let the sigma-algebra generated by all variables in the dataset be $\mathcal{F} = \sigma \{(\boldsymbol{x}_{ij}, u_{ij}, r_{T_{k}}, r_{C_{k}}, \, k = 1, \cdots, K), \, i = 1, \cdots, I, \, j = 1, 2\}$.

In a paired randomized experiment, each unit's treatment assignment is independent and equally probable:
\begin{align*}
    \mathrm{Pr}(Z_{i1} = 1 \mid \mathcal{F}) = \frac{1}{2}, \quad Z_{i2} = 1 - Z_{i1}.
\end{align*}
Thus, the joint probability of a specific assignment is:
\begin{align*}
    \mathrm{Pr}(\mathbf{Z} = \mathbf{z} \mid \mathcal{F}, \mathcal{Z}) = \frac{1}{|\mathcal{Z}|} = \frac{1}{2^I}.
\end{align*}
Under Fisher’s sharp null hypothesis, a test statistic $T_k$ for outcome $k$ is given by $T_k = t(\mathbf{Z}, \mathbf{R}_k)$. The significance level of the test follows:
\begin{align} \label{fisher_equ}
    \mathrm{Pr} \left(T_k \geq c \mid \mathcal{F}, \mathcal{Z} \right) = \frac{\left| \{ \mathbf{z} \in \mathcal{Z}: \, t(\mathbf{Z}, \mathbf{R}_k) \geq c\} \right|}{2^I}.
\end{align}

\subsection{Sensitivity Analysis}
In observational studies, treatment assignment may be influenced by unobserved confounders \citep{rosenbaum1985bias}. This results in an assignment probability $\pi_i = \mathrm{Pr}(Z_{i1} \mid \mathcal{F})$ that deviates from $\frac{1}{2}$, i.e., $\pi_i \neq \frac{1}{2}$. Sensitivity analysis assesses how varying levels of unmeasured confounding impact inferences about treatment effects. Two commonly employed approaches are sensitivity models in treatment assignments and sensitivity models in unobserved covariates. 

The sensitivity model in treatment assignments is quantified by the odds ratio of units receiving treatment in any two independent pairs across the entire population. This odd ratio is written to be bounded by the bias factor $\Gamma \geq 1$ and its reciprocal:
\begin{align} 
    \label{sensitivity_model_odd_ratio}
    \frac{1}{\Gamma} \leq \frac{\pi_{i}(1 - \pi_{j})}{\pi_{j}(1 - \pi_{i})} \leq \Gamma \text{, \, } i = 1, \cdots, I
\end{align}
with the corresponding covariates assumed to be $\boldsymbol{x}_i = \boldsymbol{x}_j$. 

Alternatively, a sensitivity model can be formulated in terms of unobserved covariates \citep{10.2307/2289789}, which affect the propensity score in a logistic model:
\begin{align}
    \label{sensitivity_model_prob}
    \mathrm{Pr}(\mathbf{Z} = \mathbf{z} \mid \mathcal{F}, \mathcal{Z}) = \prod_{i = 1}^I \frac{\mathrm{exp}(\gamma \sum_{j = 1}^2 z_{ij} u_{ij})}{\mathrm{exp}(\gamma u_{i1}) + \mathrm{exp}(\gamma u_{i2})}, 
\end{align}
where $u_{ij} \in [0,1]$, $\gamma = \log(\Gamma)$, and $\mathcal{U} = [0, 1]^{2I}$ represents the space of unobserved confounders.

These two sensitivity models—based on treatment assignment and unobserved confounders — are equivalent, see \cite{rosenbaum2002observational}. In this paper, we focus on the treatment assignment model since its parameters are more easily observed and quantified.

\subsection{Design Evaluation} 
The performance of an observational study design can be evaluated by its power of sensitivity analysis \citep{rosenbaum2004design}. Given a sensitivity model (\ref{sensitivity_model_prob}) with $\Gamma > 1$, we define the critical threshold $c_\Gamma$ as the test statistic cutoff maintaining the significance level $\alpha$. That is, $c_\Gamma$ satisfies:
\begin{align*} 
    \max_{\mathbf{u} \in \mathcal{U}} \mathrm{Pr}(T_k \geq c_\Gamma \mid \mathcal{F}) = \alpha.
\end{align*}
The power of sensitivity analysis for an alternative to Fisher's null $H_0^k$ is the probability that under that alternative, Fisher's null will be rejected even allowing for bias $\Gamma$:
\begin{align*}
    \mathrm{E}_n \left[ \mathrm{Pr}(T_k \geq c_\Gamma \mid \mathcal{F}) \right].
\end{align*}

\section{Method}
\label{s:method}
\subsection{Background}
\cite{cox1975note} introduced the data splitting approach for significance-level evaluations, in which the dataset is partitioned into two parts: one for hypothesis selection and the other for significance evaluation. Inspired by the efficiency and flexibility of the data split framework, \citep{heller2009split} proposed testing causal effects in matched pair data by randomly splitting the pairs into a planning sample and analysis sample, using the planning sample to decide which one outcome to keep and an analysis sample to test whether the treatment affects that outcome. Furthermore, \cite{bekerman2024planning} designed a method for selecting multiple outcomes in the planning sample when the sample splitting fraction is given. Building on this idea, our work addresses the following key questions: 
\begin{enumerate}
    \item The test power in the sensitivity analysis varies tremendously between different sample split fractions $\zeta$ in the simulation results of \cite{heller2009split}. How should we optimally choose $\zeta$ to maximize test power when controlling the family-wise error rate (FWER) or false discovery rate (FDR) in the presence of multiple outcomes?
    \item Following the empirical selection of the optimal sample splitting fraction, how are we going to recommend that practitioners implement a treatment effect test without requiring complicated mathematical derivations? 
\end{enumerate}

\subsection{Motivation and Contribution}
To address these challenges, we propose an \emph{Optimal Sample Split Multiple Tests} framework that provides:
\begin{enumerate}
    \item \textbf{Numerical solution for the optimal sample split fraction  $\zeta^*$:} Inspired by \cite{franklin2014plasmode} and \cite{gadbury2008evaluating}, we construct plasmode datasets to identify the optimal split fraction $\zeta^*$ across different bias levels $\Gamma$. We numerically solve for $\zeta^*$ by maximizing the test power of the treatment effect at various split fractions.
    \item \textbf{Selecting multiple outcomes to test:} The naive sample split method, which selects only the most significant outcome in both the planning and analysis stages, may be overly conservative in high-dimensional settings. To address this limitation, we introduce the \emph{Optimal Sample Split Multiple Tests} framework that leverages information from multiple affected outcomes. We integrate sample splitting with two established multiple testing procedures: 
    \begin{itemize}
        \item \emph{Optimal Sample Split FWER Approach}, which employs the \textbf{Bonferroni correction} \citep{dunn1961multiple} to control the family-wise error rate (FWER), providing conservative conclusions suitable for confirmatory analyses. Related FWER-controlling methods include those of \cite{holm1979simple, hochberg1988sharper}.
        \item \emph{Optimal Sample Split FDR Approach}, which applies the \textbf{Benjamini-Hochberg (BH) procedure} \citep{benjamini1995controlling} to control the false discovery rate (FDR), offering greater sensitivity for detecting multiple treatment effects. This approach is particularly well-suited for exploratory analyses in high-dimensional settings. Related FDR-controlling methods include those of \cite{benjamini2000adaptive, storey2002direct, genovese2004stochastic}.
    \end{itemize}
    \item \textbf{Recommendation of the sample split fraction to practitioners:} Numerical results in Section \ref{s:simulation} show that the optimal split fraction is not a single fixed value but lies within a practical range within $5\%$ error. We empirically recommend using:
    \begin{itemize}
        \item \emph{Optimal Sample Split FWER Approach} for family-wise error control with around $60\%$ of the matched pairs partitioned into the planning sample. 
        \item \emph{Optimal Sample Split FDR Approach} for false discovery rate control with around $90\%$ of the matched pairs partitioned into the planning sample.
    \end{itemize}
\end{enumerate}

\subsection{Optimal Sample Split Multiple Tests}

Given a bias factor $\Gamma \geq 1$, let $\mathcal{D} = \{(\mathbf{x}_{ij}, Z_{ij}, R_{ijk}): i = 1, \ldots, I, j = 1, 2, k = 1, \ldots, K\}$ denote the complete dataset with $I$ matched pairs and $K$ outcomes. The proposed \emph{Optimal Sample Split Multiple Tests} framework consists of two stages:
\begin{itemize}
    \item \textbf{Stage 1 (Parameter Selection):} Generate $M$ plasmode datasets $\{\mathcal{D}^{(1)}, \ldots, \mathcal{D}^{(M)}\}$ from observed control group responses $\{R_{ijk}: Z_{ij} = 0\}$ by simulating treatment effects on a randomly selected subset of outcomes. For each candidate split fraction $\zeta \in \{0.01, 0.02, \ldots, 0.99\}$, evaluate the empirical test power and select $\zeta^*$ that maximizes this power:
    \begin{itemize}
        \item \textbf{Step 1 (Sample Splitting):} For each plasmode dataset $\mathcal{D}^{(m)}$ and split fraction $\zeta$, randomly partition the $I$ matched pairs into a planning sample $\mathcal{D}_{\text{plan}}^{(m)}$ of size $\lfloor (1-\zeta)I \rfloor$ and an analysis sample $\mathcal{D}_{\text{anal}}^{(m)}$ of size $\lceil \zeta I \rceil$.
        \item \textbf{Step 2 (Hypothesis Planning):} Test each outcome $k \in \{1, \ldots, K\}$ under $H_0^k: \delta_{i} = 0$ on the planning sample $\mathcal{D}_{\text{plan}}^{(m)}$ using test statistic $t(\mathbf{Z}_{\text{plan}}, \mathbf{R}_{k,\text{plan}})$ to obtain $p$-values $p_1^{\text{plan}}, \ldots, p_K^{\text{plan}}$. Select promising outcomes for further testing:
        \begin{itemize}
            \item \emph{FWER approach:} Apply Bonferroni correction at level $\alpha/K$ to define the selected set $\mathcal{S}^{(m)}(\zeta) = \{k: p_k^{\text{plan}} \leq \alpha/K\}$ with $s^{(m)}(\zeta) = |\mathcal{S}^{(m)}(\zeta)|$.
            
            \item \emph{FDR approach:} Apply Benjamini-Hochberg procedure to select outcomes for analysis. Order the planning sample $p$-values as $p_{(1)}^{\text{plan}} \leq \cdots \leq p_{(K)}^{\text{plan}}$ with corresponding outcome indices $k_{(1)}, \ldots, k_{(K)}$. Find the cutoff $\ell^* = \max\{\ell: p_{(\ell)}^{\text{plan}} \leq \frac{\ell}{K}\alpha\}$ and define the selected set $\mathcal{G}^{(m)}(\zeta) = \{k_{(1)}, \ldots, k_{(\ell^*)}\}$ of size $\ell^* = |\mathcal{G}^{(m)}(\zeta)|$ to be used in Step 3 for analysis.
        \end{itemize}
        \item \textbf{Step 3 (Significance Analysis):} Conduct analysis step hypothesis testing on the analysis sample $\mathcal{D}_{\text{anal}}^{(m)}$ for the selected outcomes:
        \begin{itemize}
            \item \emph{FWER approach:} For each outcome $k \in \mathcal{S}^{(m)}(\zeta)$, compute the analysis sample $p$-value $p_k^{\text{anal}}$ using test statistic $t(\mathbf{Z}_{\text{anal}}, \mathbf{R}_{k,\text{anal}})$. Reject $H_0^k$ if $p_k^{\text{anal}} \leq \alpha/s^{(m)}(\zeta)$ to control FWER at level $\alpha$.
            \item \emph{FDR approach:} We can use the Bonferroni procedure to reject $H_0^k$. For each outcome $k \in \mathcal{G}^{(m)}(\zeta)$, compute the analysis sample $p$-value $p_k^{\text{anal}}$. Apply the Benjamini-Hochberg procedure to these $\ell^*$ analysis sample $p$-values: order them as $p_{(1)}^{\text{anal}} \leq \cdots \leq p_{(\ell^*)}^{\text{anal}}$ with corresponding outcome indices, find $\ell^{**} = \max\{j: p_{(j)}^{\text{anal}} \leq \frac{j}{\ell^*}\alpha\}$, and reject $H_0^k$ for all outcomes corresponding to $p_{(1)}^{\text{anal}}, \ldots, p_{(\ell^{**})}^{\text{anal}}$. This controls FDR at level $\alpha$ for the selected outcomes. We could also use the Holm procedure to increase the power \citep{holm1979simple}, a simple sequentially rejective multiple test procedure. 
        \end{itemize}
        \item \textbf{Step 4 (Power Evaluation):} For each candidate split fraction $\zeta \in \{0.01, 0.02, \ldots, 0.99\}$, repeat Steps 1-3 over $M$ plasmode replications. In each plasmode replication $m$, let $\mathcal{H}_1^{(m)} = \{k: \delta_{ik}^{(m)} \neq 0\}$ denote the set of truly affected outcomes with $|\mathcal{H}_1^{(m)}| = \lfloor \eta K \rfloor$ as assumed, where $\eta \in (0, 1)$ is the proportion of affected outcomes. Let $\mathcal{R}^{(m)}(\zeta)$ denote the set of rejected hypotheses in replication $m$ at split fraction $\zeta$. The empirical power function is:
        \begin{align}
            \hat{\pi}(\zeta) = \frac{1}{M} \sum_{m=1}^{M} \frac{|\mathcal{R}^{(m)}(\zeta) \cap \mathcal{H}_1^{(m)}|}{|\mathcal{H}_1^{(m)}|}. \label{eq:power_function}
        \end{align}
        The optimal split fraction maximizing this empirical power is:
        \begin{align}
            \zeta^* = \arg\max_{\zeta \in (0,1)} \hat{\pi}(\zeta). \label{eq:optimal_zeta}
        \end{align}
    \end{itemize}
    \item \textbf{Stage 2 (Treatment Effect Tests):} Apply the optimal split fraction $\zeta^*$ from (\ref{eq:optimal_zeta}) to the original dataset $\mathcal{D}$. Partition the matched pairs into planning sample $\mathcal{D}_{\text{plan}}$ and analysis sample $\mathcal{D}_{\text{anal}}$. Conduct hypothesis planning (Step 2) on $\mathcal{D}_{\text{plan}}$ and significance analysis (Step 3) on $\mathcal{D}_{\text{anal}}$ using the observed data. Let $\mathcal{R}_{\text{final}}$ denote the set of rejected hypotheses, representing outcomes significantly affected by treatment at bias level $\Gamma$.
\end{itemize}

\paragraph{Implementation Details}
To evaluate the power function (\ref{eq:power_function}) and identify $\zeta^*$, we generate $M$ plasmode datasets from observed control responses \citep{franklin2014plasmode, gadbury2008evaluating}. For each plasmode replication $m$, we randomly select $\lfloor \eta K \rfloor$ outcomes to be affected, where $\eta \in [0.075, 0.30]$ following \cite{korthauer2019practical}. Treatment effects are drawn as $\delta_{ik}^{(m)} \sim \text{Uniform}[a \cdot \sigma_k, b \cdot \sigma_k]$ with $(a, b)$ chosen for small to medium effect sizes - around $0.2$ to $0.5$ according to \citep{stuart2011matchit, sawilowsky2009new}, and outcomes are constructed as $R_{ijk}^{(m)} = r_{C_{ik}} + Z_{ij} \delta_{ik}^{(m)}$, where $r_{C_{ik}}$ is the outcome on the unit receives control. Since the resulting power function is typically flat near its maximum \citep{heller2009split}, we recommend using any $\zeta$ satisfying $\hat{\pi}(\zeta) \geq 0.95 \cdot \hat{\pi}(\zeta^*)$, which yields an interval of near-optimal fractions as demonstrated in the result tables of Section \ref{s:simulation}.

\subsection{Example: Wilcoxon’s Signed Rank Statistic}
Various statistical tests can be applied for treatment effect evaluation \cite{rosenbaum2011new, zhao2018multiple, shauly2020exact, howard2021uniform}. In this study, we use Wilcoxon's signed rank statistics \citep{10.2307/3001968}. The test statistic is widely used due to its asymptotic normality. The test statistic is:
\begin{align*}
    t(\mathbf{Z}, \mathbf{R}_k) = \sum_{i = 1}^I \mathrm{sgn}\{(R_{i1} - R_{i2})(Z_{i1} - Z_{i2})\} \cdot \mathrm{rank}(|R_{i1} - R_{i2}|),
\end{align*}
where $\mathrm{rank}(\cdot)$ is the rank of observations from small to large with average ranks for ties, and $\mathrm{sgn}(w)$ is a function of sign with $\mathrm{sgn}(w) = 0, \frac{1}{2}, 1$ as $w < 0$, $w = 0$, and $w > 0$. 

\subsection{Asymptotic Properties of Wilcoxon's Signed Rank Statistic}
\label{s:asymptotic}
We examine the asymptotic behavior of Wilcoxon's signed rank test in the context of sensitivity analysis. Specifically, we analyze the critical value, test power in randomized experiments, and test power under unmeasured confounding.

Suppose $T_k$ is the Wilcoxon signed rank test statistic, denoted as $W_k = w(\mathbf{Z}, \mathbf{R}_k)$ and the bias factor is $\Gamma > 1$ with a large sample size $I$, the critical value $c_\Gamma$ at the significance level $\alpha$ is defined by:
\begin{align*} 
    \max_{\mathbf{u} \in \mathcal{U}} \mathrm{Pr}(W_k \geq c_\Gamma \mid \mathcal{F}) = \alpha.
\end{align*}
The \textbf{design sensitivity} is defined as the threshold $\widetilde{\Gamma}$ at which the power of sensitivity analysis goes from converging to 1 to converging to 0 as I converges to infinity, i.e.,
\begin{align*}
    \mathrm{Pr}(W_k \geq c_\Gamma \mid \mathcal{F}) \to 
    \begin{cases}
        1, & \text{if } \Gamma < \widetilde{\Gamma}, \\
        0, & \text{if } \Gamma > \widetilde{\Gamma},
    \end{cases}
\end{align*}
The details of the asymptotic behaviors are explained in the supplemental materials. 

\section{Simulation}
\label{s:simulation}
\subsection{Simulation Methods}
In this simulation study, we compare our proposed method against existing approaches on observational datasets. The candidate methods include:
\begin{itemize}
    \item \textbf{Bonferroni Correction:} A traditional method for multiple outcome tests, controlling family-wise error rate (FWER) by adjusting the significance level to $\alpha / K$, where $K$ is the number of hypotheses. 
    \item \textbf{Naive (Traditional) Sample Split Test:} This method selects the outcome with the most significant $p$-value in the planning step and then tests its hypothesis in the analysis step. 
    \item \textbf{Optimal Sample Split FWER Approach:} An extension of sample split that incorporates multiple hypothesis testing, selecting one or more significant outcomes in the planning step, and conducting multiple tests in the analysis step.    
    \item \textbf{Optimal Sample Split FDR Approach:} This method uses the Benjamini-Hochberg procedure to select outcomes in the planning step, then applies the Benjamini-Hochberg procedure again to those selected outcomes in the analysis step, controlling FDR at level $\alpha$ for the final rejections.
\end{itemize}

\subsection{Data Generating Process}
We generate datasets containing covariates $\mathbf{X}$, treatment assignments $\mathbf{Z}$, and corresponding outcomes $\mathbf{R}$. Full matching is performed using the \texttt{MatchIt} package \cite{stuart2011matchit} to create matched-pair datasets suitable for our methods.

Each dataset consists of $N = 5000$ observations with $D = 5$ covariates. We vary the number of outcomes across $K \in \{10, 100, 500, 1000\}$ to compare performance under both low and high-dimensional settings. We quantify the sensitivity of the results of our hypothesis test using the bias factor $\Gamma$. In our simulation settings, we consider $\Gamma \in \{1, 1.25, 1.5, 1.75, 2\}$. The bias factor $\Gamma$ measures how much deviation from random treatment assignment our test can tolerate while still maintaining its conclusions. Specifically, $\Gamma = 1$ corresponds to a randomized experiment with no unmeasured confounding, while larger values of $\Gamma$ indicate a greater potential bias in treatment assignment due to unmeasured confounders. 

The treatment assignment is generated according to the sensitivity model:
$Z \sim \text{Bernoulli} \left( \frac{\Gamma}{1 + \Gamma^2} \right)$.
Covariates are drawn from a uniform distribution $\mathbf{X} \sim \text{Uniform}[0,5]^D$, and the unmeasured confounder follows $U \sim \mathcal{N}(0,1)$. The treatment effect on affected outcomes is modeled as $\tau \sim \mathcal{N}(1,1)$. We assume that $\eta = 10\%$ of the total outcomes are affected. The outcome model is given by:
\begin{align*}
    \text{Affected outcomes:} & \quad R(\mathbf{X}, Z) = \mathbf{X} \boldsymbol{\alpha} + \tau Z + \epsilon, \\
    \text{Unaffected outcomes:} & \quad R(\mathbf{X}, Z) = \mathbf{X} \boldsymbol{\alpha} + \epsilon,
\end{align*}
where each entry $\alpha_i$ in $\boldsymbol{\alpha} = (\alpha_1, \dots, \alpha_D)$ is sampled from $\mathcal{N}(1,1)$. After generating the dataset, full matching is performed, and subsets are drawn to create matched-pair datasets of size $I \in \{100, 200, 500, 1000\}$.

Since high-dimensional settings typically exhibit correlated features, we generate correlated outcomes to reflect realistic scenarios.

\subsection{Simulation Implementation}
For each matched-pair dataset generated across different values of $\Gamma$, $I$, and $K$, we conduct hypothesis testing on the treated-minus-control difference for each outcome. Define the treated-minus-control difference for the $k^{\text{th}}$ outcome as:
\[V_{ik} = (R_{i1k} - R_{i2k})(Z_{i1} - Z_{i2}),\]
and let $\mathbf{V}_k = [V_{ik}]_{i=1}^{I}$ represent all matched pairs. The hypotheses are:
\begin{align*}
    H_{0}^k: & \quad \mathbf{V}_k = 0, \\
    H_{1}^k: & \quad \mathbf{V}_k \neq 0.
\end{align*}
Each scenario is replicated $100$ times with a significance level of $\alpha = 0.05$.

We present the simulation results for $K = 1000$ in Table \ref{tab:sim-k-1000} and leave the results of $K = 10, 100, 500$ to Section C of the Supplemental Materials. For each method, the first row reports test power, while the second row (for sample splitting methods) indicates the recommended set of splitting fractions $\zeta$. 

The reported test powers for the sample split methods correspond to the optimal fractions for each setting. 
It is noteworthy that in some simulation scenarios, the optimal split fraction is not a single value but rather an interval of fractions that achieves the maximum power. 
To provide further insight, we report the recommended sample split fraction sets, which include not only the fraction(s) yielding the maximum test power but also those achieving at least $95\%$ of this maximum. 

Overall, the simulation results indicate that the Optimal Sample Split FDR Approach consistently achieves higher power than the baseline methods - Bonferroni Correction and Naive (Traditional) Sample Split Test - across all values of sample size $I$, number of outcomes $K$, and design sensitivity $\Gamma$. 
In contrast, the Optimal Sample Split FWER pproach generally achieves higher power than the baseline methods and, in some instances, performs comparably. 

Key observations from the simulation results include:
\begin{enumerate}
    \item \textbf{Effectiveness in High Bias and High Dimension:} The Optimal Sample Split FWER Approach generally achieves higher power the baseline familywise error rate (FWER)-controlling methods (Bonferroni Correction and Naive Sample Split Test) in terms of test power. The performance gains are especially marked when the bias factor is high ($\Gamma > 1$) or when the outcome space is high-dimensional under small sample size cases. These results underscore the superior performance of our sample splitting approach in challenging settings where traditional methods tend to falter. The numerical findings are consistent with the theoretical analysis, which emphasizes the advantages of sample splitting in high-bias contexts.
    \item \textbf{Performance Under Low Bias}: When the bias factor is low ($\Gamma = 1$), the Optimal Sample Split FWER Approach performs comparably to the Bonferroni Correction in most settings, and both achieves higher power the Naive Sample Split Test. This indicates that our sample splitting approach with proper fraction selection effectively addresses the shortcomings of the naive approach in unbiased settings, echoing observations made in \cite{heller2009split}.
    \item \textbf{Exploratory Analysis for High-dimensional Settings}: The Optimal Sample Split FDR Approach controls the false discovery rate (FDR) at level $\alpha$ rather than the familywise error rate (FWER), providing a more liberal and powerful approach for detecting multiple treatment effects. This method demonstrates consistently high power across all simulation settings, particularly excelling in high-dimensional outcome spaces and maintaining strong performance even under substantial bias ($\Gamma > 1$). This makes the approach particularly well-suited for exploratory research in high-dimensional settings where the goal is to identify multiple potentially affected outcomes across numerous pathways.
    \item \textbf{Necessity of selecting an optimal sample split fraction:} Test power varies considerably across splitting fractions, with a range of fractions achieving near-optimal performance. Practitioners are advised to use $\zeta \approx 90\%$ for rank methods (FDR control) and $\zeta \approx 60\%$ for selection methods (FWER control).
    \item \textbf{Rapid Convergence with Increasing Sample Size:} As the sample size $I$ increases, both the Optimal Sample Split FWER Approach and the Optimal Sample Split FDR Approach converge to their theoretical asymptotic test power more rapidly than the baseline methods. This finding confirms the fast convergence rate of the proposed method.  
\end{enumerate}

\begin{table}[]
\centering
\begin{tabular}{llllll}
\hline
      & & \multicolumn{4}{c}{I}\\
      & & 100    & 200   & 500   & 1000\\ 
      \cline{3-6} 
      Gamma & & & & &\\ 
      \hline
      1 & Bonferroni                & 0.186 & 0.350 & 0.500 & 0.596\\
      & Naive split                 & 0.01 &         0.01 &         0.01 &         0.01 \\
      &                             & [0.09, 0.89] & [0.04, 0.99] & [0.02, 0.97] & [0.01, 0.99] \\
      & Optimal split selection  & 0.167 &       0.3066 &       0.4739 &       0.5653 \\
      &                             & [0.6, 0.8] & [0.58, 0.76] &  [0.49, 0.8] & [0.46, 0.83] \\
      & Optimal split rank   & 0.3989 &       0.5343 &       0.6357 &       0.7083 \\
      &                             & [0.91, 0.97] &  [0.9, 0.99] & [0.84, 0.99] &  [0.8, 0.99] \\
      \hline
      1.25 & Bonferroni             & 0.136 & 0.264 & 0.404 & 0.517\\
      & Naive split                 & 0.01 &         0.01 &         0.01 &         0.01 \\
      &                             & [0.15, 0.84] & [0.08, 0.94] & [0.03, 0.97] & [0.02, 0.99] \\
      & Optimal split selection  & 0.1359 &       0.2469 &       0.3839 &       0.4926 \\
      &                             & [0.59, 0.77] & [0.57, 0.77] &  [0.53, 0.8] &  [0.52, 0.8] \\
      & Optimal split rank   & 0.3096 &        0.438 &       0.5568 &       0.6304 \\
      &                             & [0.92, 0.98] & [0.93, 0.99] & [0.94, 0.99] & [0.89, 0.99] \\
      \hline
      1.5 & Bonferroni              & 0.004 & 0.030 & 0.104 & 0.162\\
      & Naive split                 & 0.0075 &       0.0092 &         0.01 &         0.01 \\
      &                             & [0.78, 0.8] &  [0.5, 0.87] & [0.21, 0.95] &  [0.1, 0.97] \\
      & Optimal split selection  & 0.0148 &       0.0385 &       0.1057 &        0.161 \\
      &                             & [0.7, 0.7] & [0.57, 0.76] & [0.57, 0.73] & [0.48, 0.77] \\
      & Optimal split rank   & 0.0555 &       0.0992 &       0.1774 &       0.2203 \\
      &                             & [0.95, 0.96] & [0.97, 0.99] & [0.98, 0.99] & [0.98, 0.99] \\
      \hline
      1.75 & Bonferroni             & 0.029 & 0.043 & 0.096 & 0.156\\
      & Naive split                 & 0.01 &         0.01 &         0.01 &         0.01 \\
      &                             & [0.24, 0.81] & [0.12, 0.93] & [0.05, 0.97] & [0.03, 0.98] \\
      & Optimal split selection  & 0.0334 &       0.0523 &       0.0989 &       0.1559 \\
      &                             & [0.53, 0.69] &  [0.6, 0.77] & [0.56, 0.74] & [0.55, 0.74] \\
      & Optimal split rank   & 0.0626 &       0.1099 &       0.1708 &       0.2157 \\
      &                             & [0.95, 0.98] & [0.98, 0.99] & [0.98, 0.99] & [0.98, 0.99] \\
      \hline
      2 & Bonferroni                & 0.010 & 0.017 & 0.049 & 0.061\\
      & Naive split                 & 0.01 &         0.01 &         0.01 &         0.01 \\
      &                             & [0.27, 0.74] & [0.14, 0.89] & [0.05, 0.96] & [0.03, 0.97] \\
      & Optimal split selection  & 0.0144 &       0.0262 &       0.0528 &       0.0661 \\
      &                             & [0.48, 0.78] &  [0.5, 0.69] &   [0.5, 0.7] & [0.48, 0.74] \\
      & Optimal split rank   & 0.0235 &       0.0411 &       0.0688 &       0.0829 \\
      &                             & [0.95, 0.98] & [0.98, 0.99] & [0.99, 0.99] & [0.99, 0.99] \\
      \hline
\end{tabular}
\caption{Simulation result, when the number of outcomes $K = 1000$: The table presents the test powers and the corresponding recommended sample split fraction, set at the given number of outcomes $K = 1000$ for each method under various bias factors and sample sizes. The simulation is replicated $1000$ times with the significance level to be $\alpha = 0.05$. The first row for each simulation setting presents the test power, while the brackets in the second row (for the split methods) indicate the recommended set for the split fraction $\zeta$}.
\label{tab:sim-k-1000}
\end{table}

\subsection{Effectiveness and Robustness of Optimal Sample Split with Multiple Tests}
To further illustrate the effectiveness and robustness of the \emph{Optimal Sample Split with Multiple Tests}, we visualize the test power under different sample split fractions.  
For this analysis, we consider $K = 10$ outcomes and a sample size of $I = 200$ under the bias factor $\Gamma = 1$, with $1000$ replications.  
\begin{figure}
    \centering
    \includegraphics[width=0.5\linewidth]{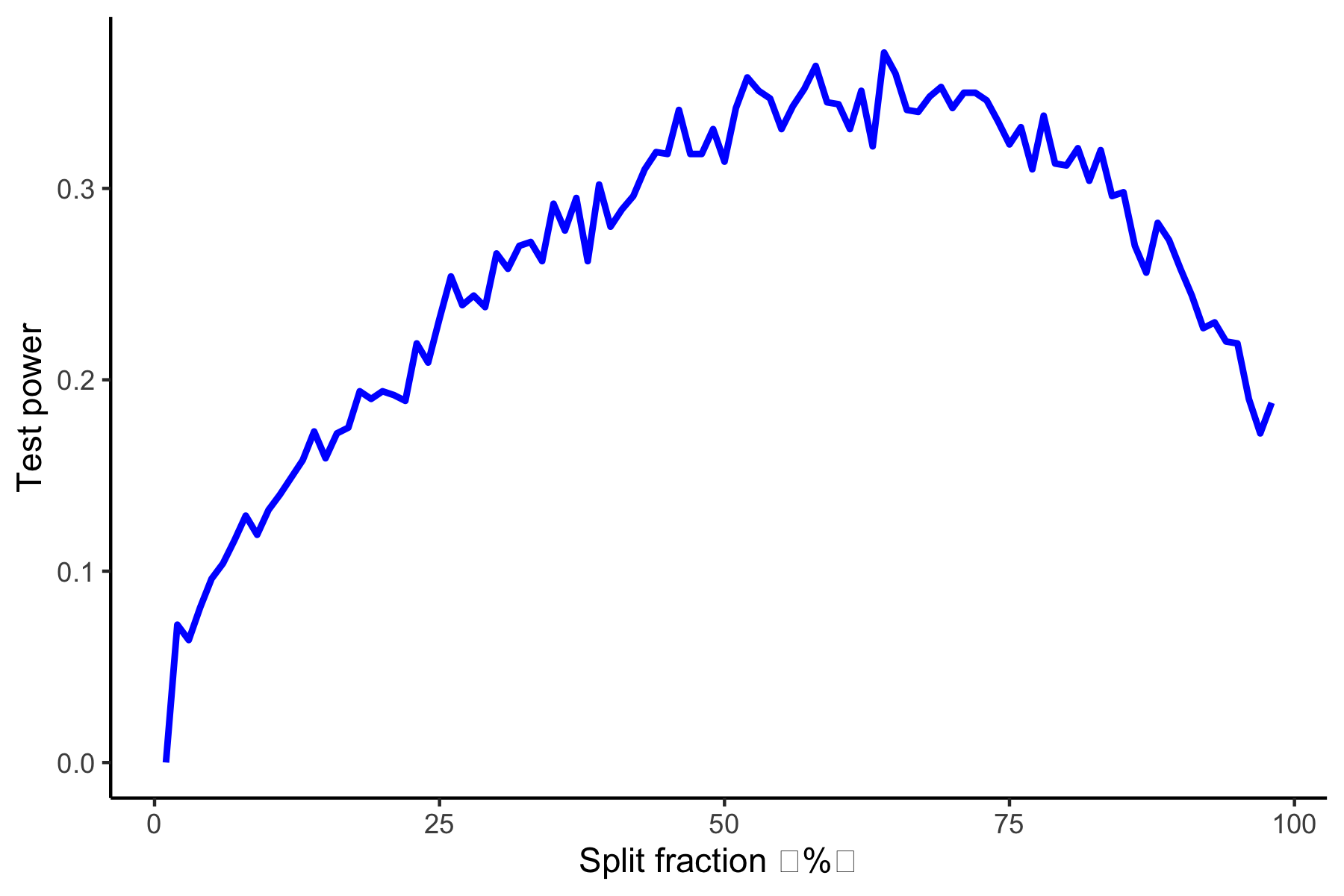}
    \caption{Test power vs. sample split fraction under multiple hypothesis tests for outcomes $K = 10$, sample size $I = 200$, and bias factor $\Gamma = 1$ by selection with $1000$ replications.}
    \label{fig:ex-power}
\end{figure}

\paragraph{Effectiveness of the Optimal Sample Split Method}
As shown in Figure~\ref{fig:ex-power}, the \emph{Optimal Sample Split with Multiple Tests} significantly improves the power of detecting treatment effects compared to naive sample splitting methods.  
The test power follows an \textbf{``upside-down U''} shape, where different sample split fractions yield varying levels of power.  
Notably, when the sample split fraction is appropriately chosen, the test power reaches a higher peak than the naive sample split method, confirming the effectiveness of our approach in maximizing detection sensitivity.  
In contrast, a naive sample split method often yields suboptimal power, potentially missing significant treatment effects.

\paragraph{Robustness of the Optimal Sample Split Method}
Beyond its effectiveness, the proposed method also exhibits robustness in its performance.  
Within a moderate range of sample split fractions, the test power remains consistently close to its peak.  
Even if the exact optimal fraction is not chosen, the test power fluctuates only negligibly from the maximum, demonstrating that the method is not highly sensitive to slight deviations in the split fraction.  
This robustness ensures that practitioners can achieve reliable results without requiring precise tuning of the split fraction.  
Compared to the naive sample split method, the \emph{Optimal Sample Split with Multiple Tests} provides a stable and reliable framework for detecting treatment effects in observational studies.

\section{Application}
\label{s:application}
According to \cite{braun2020influence}, the effect of second-hand smoke on children's health can be observed from various aspects. Researchers have studied how secondhand smoke affects blood lead levels, but a comprehensive quantitative analysis of multiple health effects is still lacking. In this work, we apply our proposed \textit{Optimal Sample Split FDR Approach} to examine the effect of secondhand smoke under different bias factor levels. 
The reason we chose the FDR approach rather than the FWER is that we do not want to miss too many outcomes that secondhand smoke affects and  would regard it as a success if say we reported 20 outcomes that secondhand smoke affected and 19 were corrected and 1 was wrong.
We conduct a sensitivity analysis using data from the US National Health and Nutrition Examination Survey (NHANES) 2017-2018 at \cite{nhanes2017overview}. 

Following \cite{zhang2020calibrated}, a child is classified as being exposed to a high level of second-hand smoke if the cotinine level is greater than or equal to $0.563 \, \text{ng/ml}$, i.e., $\text{treatment} = 1$, and otherwise classified as not exposed, i.e., $\text{treatment} = 0$ and perform full matching based on education, gender, race/ethnicity, age, poverty level, number of rooms in the house, and family size. This results in 154 pairs matched and the corresponding box plot for the matched nonbinary covariates is presented in the supplement materials.

In this application study, we conduct a sensitivity analysis to determine the effect of second-hand smoke exposure on children's laboratory outcomes under different bias factors. After matching, we analyze the effects of second-hand smoke on $76$ laboratory outcomes for bias factors $\Gamma = \{1, 1.25, 1.5, 1.75, 2\}$. To numerically solve the optimal fraction for sample splitting, we generate $1000$ plasmode datasets for simulation. We assume a ratio of outcomes affected by the treatment of $\eta = 10\%$ and randomly assign treatment effects when generating the plasmode datasets. Since most of the treatment-minus-control differences for each laboratory outcome are assumed to be subtle, we set the effect sizes to be drawn within a small range, $\mathrm{Unif}(0.05 \cdot \text{std}, 0.20 \cdot \text{std})$, according to \cite{sawilowsky2009new}, which interprets and suggests effect sizes. Although the various values of hyperparameters when generating the plasmode datasets, the solutions of the optimal sample split fraction are within an interval, $[0.84, 0.98]$. The solution demonstrates the robustness of the proposed optimal sample split method regardless of the selection of hyperparameters in the plasmode datasets generation. We conduct the subsequent analysis by splitting the analysis sample of $90\%$, which is a commonly used sample split ratio within the solution of the optimal sample split fraction interval, and it is also recommended by the simulation results. In the synthetic plasmode dataset to select the optimal sample split fractions, the ratio of the affected outcomes $\eta$ and effect size are free to be chosen to match different application cases. 

Tables~\ref{tab:application_1} and \ref{tab:application_2} report the results of whether each outcome is significantly affected by second-hand smoke exposure at different bias factor levels. A value of $1$ indicates that the outcome is significantly affected, while $0$ indicates no significant effect. A detailed description of each outcome is available at \cite{CDC_NHANES_2017_2018}. 

The laboratory outcomes identified as significantly affected by second-hand smoke exposure include: \textit{Creatinine, total Cholesterol, red cell distribution width, HS C-reactive protein, Alkaline Phosphatase (ALP), Blood Urea Nitrogen, Creatine Phosphokinase (CPK), refrigerated serum Creatinine, Lactate Dehydrogenase (LDH), Cholesterol, Uric acid, Transferrin receptor, and gamma-tocopherol}. 

For comparison, we also report results from the \textit{Bonferroni correction} method applied to the full dataset. All outcomes identified as significant under the Bonferroni correction are also detected using our proposed \textit{Optimal Sample Split FDR Approach}. These outcomes are marked with an asterisk (*) in Tables~\ref{tab:application_1} and \ref{tab:application_2}. When the bias factor $\Gamma$ is large, the Bonferroni correction fails to detect the treatment effect, whereas our method continues to identify affected outcomes. This highlights the advantage of our approach in high-bias settings.

The findings from our \textit{Optimal Sample Split with Multiple Tests} align with previous medical studies, including \cite{mannino2003second}, \cite{smith2003effect}, and \cite{braun2020influence}, in terms of the outcomes identified as affected by second-hand smoke exposure, further supporting the validity of our approach.

\begin{table}[]
    \centering
    \begin{tabular}{c c c c c c c }
         \hline
         ~ & ~ & \multicolumn{5}{c}{Bias factor $\Gamma$} \\
         \multicolumn{2}{c}{Outcomes} & 1 & 1.25 & 1.5   & 1.75 & 2\\
         \hline
         LBXHCT & Hematocrit $\%$ & 0 & 0 & 0 & 0 & 0\\
         URXUMA & Albumin, urine (ug/mL) & $1$ & 0 & 0 & 0 & 0\\
         URXUCR & Creatinine, urine (mg/dL) & $1^*$ & $1^*$ & 0 & 0 & 0\\
         LBXTC & Total Cholesterol (mg/dL) & $1^*$ & 0 & 0 & 0 & 0\\
         LBXWBCSI & White blood cell count (1000 cells/uL) & 0 & 0 & 0 & 0 & 0\\
         LBDLYMNO & Lymphocyte number (1000 cells/uL) & 0 & 0 & 0 & 0 & 0\\
         LBDLYMNO & Lymphocyte number (1000 cells/uL) & 0 & 0 & 0 & 0 & 0\\
         LBDNENO & Segmented neutrophils num (1000 cell/uL) & 0 & 0 & 0 & 0 & 0\\
         LBDEONO & Eosinophils number (1000 cells/uL) & 0 & 0 & 0 & 0 & 0\\
         LBDBANO & Basophils number (1000 cells/uL) & 0 & 0 & 0 & 0 & 0\\
         LBXRBCSI & Red blood cell count (million cells/uL) & 0 & 0 & 0 & 0 & 0\\
         LBXHGB & Hemoglobin (g/dL) & 0 & 0 & 0 & 0 & 0\\
         LBXMCVSI & Mean cell volume (fL) & 0 & 0 & 0 & 0 & 0\\
         LBXMC & Mean Cell Hgb Conc. (g/dL) & 0 & 0 & 0 & 0 & 0\\
         LBXRDW & Red cell distribution width $\%$ & $1^*$ & 1 & 1 & 0 & 0\\
         LBXPLTSI & Platelet count (1000 cells/uL) & 0 & 0 & 0 & 0 & 0\\
         LBXMPSI & Mean platelet volume (fL) & 0 & 0 & 0 & 0 & 0\\
         LBXNRBC & Nucleated red blood cells & 0 & 0 & 0 & 0 & 0\\
         LBXFER & Ferritin (ng/mL) & 0 & 0 & 0 & 0 & 0\\
         LBDRFO & RBC folate (ng/mL) & 0 & 0 & 0 & 0 & 0\\
         LBDFOTSI & Serum total folate (nmol/L) & 0 & 0 & 0 & 0 & 0\\ 
         LBXGH & Glycohemoglobin ($\%$) & 0 & 0 & 0 & 0 & 0\\
         LBXHSCRP & HS C-Reactive Protein (mg/L) & $1^*$ & $1^*$ & $1^*$ & 1 & 0\\
         LBDTIB & Total Iron Binding Capacity TIBC (ug/dL) & 0 & 0 & 0 & 0 & 0\\
         LBXBPB & Blood lead (ug/dL) & 0 & 0 & 0 & 0 & 0\\
         LBXBCD & Blood cadmium (ug/L) & 0 & 0 & 0 & 0 & 0 \\
         LBXTHG & Blood mercury, total (ug/L) & 0 & 0 & 0 & 0 & 0\\
         LBXBSE & Blood selenium (ug/L) & 0 & 0 & 0 & 0 & 0\\
         LBXBMN & Blood manganese (ug/L) & 0 & 0 & 0 & 0 & 0\\
         LBXIHG & Mercury, inorganic (ug/L) & 0 & 0 & 0 & 0 & 0\\
         LBXBGE & Mercury, ethyl (ug/L) & 0 & 0 & 0 & 0 & 0\\
         LBXBGM & Mercury, methyl (ug/L) & 0 & 0 & 0 & 0 & 0\\
         LBXSATSI & Alanine Aminotransferase (ALT) (U/L) & 0 & 0 & 0 & 0 & 0\\
         LBXSAL & Albumin, refrigerated serum (g/dL) & 0 & 0 & 0 & 0 & 0\\
         LBXSAPSI & Alkaline Phosphatase (ALP) (IU/L) & $1^*$ & 0 & 0 & 0 & 0\\
         LBXSC3SI & Bicarbonate (mmol/L) & 0 & 0 & 0 & 0 & 0\\
         LBXSBU & Blood Urea Nitrogen (mg/dL) & $1^*$ & $1^*$ & $1^*$ & 0 & 0\\
         LBXSCLSI & Chloride (mmol/L) & 0 & 0 & 0 & 0 & 0\\
         LBXSCK & Creatine Phosphokinase (CPK) (IU/L) & $1^*$ & 0 & 0 & 0 & 0\\
         LBXSCR & Creatinine, refrigerated serum (mg/dL) & $1^*$ & 0 & 0 & 0 & 0 \\
         LBXSGB & Globulin (g/dL) & 0 & 0 & 0 & 0 & 0\\
         LBXSGL & Glucose, refrigerated serum (mg/dL) & 0 & 0 & 0 & 0 & 0\\
         LBXSGTSI & Gamma Glutamyl Transferase (GGT) (IU/L) & 0 & 0 & 0 & 0 & 0\\
         LBXSIR & Iron, refrigerated serum (ug/dL) & 0 & 0 & 0 & 0 & 0\\
         \hline   
    \end{tabular}
    \caption{Application on NHANES 2017-2018 (1): `1' indicates testable and `0' indicates not testable by the Optimal Sample Split with Multiple Tests. For comparison, `*' indicates testable by Bonferroni Correction.}
    \label{tab:application_1}
\end{table}

\begin{table}[]
    \centering
    \begin{tabular}{c c c c c c c}
         \hline
         ~ & ~ & \multicolumn{5}{c}{Bias factor $\Gamma$} \\
         \multicolumn{2}{c}{Outcomes} & 1 & 1.25 & 1.5   & 1.75 & 2\\
         \hline
         LBXSLDSI & Lactate Dehydrogenase (LDH) (IU/L) & $1^*$ & $1^*$ & 1 & 0 & 0\\
         LBXSOSSI & Osmolality (mmol/Kg) & 0 & 0 & 0 & 0 & 0\\
         LBXSPH & Phosphorus (mg/dL) & 0 & 0 & 0 & 0 & 0\\
         LBDSPHSI & Phosphorus (mmol/L) & 0 & 0 & 0 & 0 & 0\\
         LBXSKSI & Potassium (mmol/L) & 0 & 0 & 0 & 0 & 0\\
         LBXSNASI & Sodium (mmol/L) & 0 & 0 & 0 & 0 & 0\\
         LBXSTB & Total Bilirubin (mg/dL) & 0 & 0 & 0 & 0 & 0\\
         LBXSCA & Total Calcium (mg/dL) & 0 & 0 & 0 & 0 & 0\\
         LBXSCH & Cholesterol, refrigerated serum (mg/dL) & $1^*$ & 0 & 0 & 0 & 0 \\
         LBXSTP & Total Protein (g/dL) & 0 & 0 & 0 & 0 & 0\\
         LBXSTR & Triglycerides, refrig serum (mg/dL) & 0 & 0 & 0 & 0 & 0\\
         LBXSUA & Uric acid (mg/dL) & $1^*$ & $1^*$ & $1^*$ & 0 & 0\\
         LBXTFR & Transferrin receptor (mg/L) & $1^*$ & $1^*$ & $1^*$ & 1 & 1\\
         LBXALC & alpha-carotene (ug/dL) & 0 & 0 & 0 & 0 & 0\\
         LBXARY & alpha-crypotoxanthin (ug/dL) & 0 & 0 & 0 & 0 & 0\\
         LBXBEC & trans-beta-carotene (ug/dL) & 0 & 0 & 0 & 0 & 0\\
         LBXCBC & cis-beta-carotene (ug/dL) & 0 & 0 & 0 & 0 & 0\\
         LBXCRY & beta-cryptoxanthin (ug/dL) & 0 & 0 & 0 & 0 & 0\\
         LBXGTC & gamma-tocopherol (ug/dL) & $1^*$ & $1^*$ & $1^*$ & $1^*$ & $1^*$\\
         LBXLUZ & Lutein and zeaxanthin (ug/dL) & 0 & 0 & 0 & 0 & 0\\
         LBXLYC & trans-lycopene (ug/dL) & 0 & 0 & 0 & 0 & 0\\
         LBXRPL & Retinyl palmitate (ug/dL) & 0 & 0 & 0 & 0 & 0\\
         LBXRST & Retinyl stearate (ug/dL) & 0 & 0 & 0 & 0 & 0\\
         LBXLCC & Total Lycopene (ug/dL) & 0 & 0 & 0 & 0 & 0\\
         LBXVIA & Retinol (ug/dL) & 0 & 0 & 0 & 0 & 0\\
         LBXVIE & alpha-tocopherol (ug/dL) & 0 & 0 & 0 & 0 & 0\\
         LBXVIC & Vitamin C (mg/dL) & 0 & 0 & 0 & 0 & 0\\
         LBXVIDMS & 25OHD2+25OHD3 (nmol/L) & 0 & 0 & 0 & 0 & 0\\
         LBXVD2MS & 25OHD2 (nmol/L) & 0 & 0 & 0 & 0 & 0\\
         LBXVD3MS & 25OHD3 (nmol/L) & 0 & 0 & 0 & 0 & 0\\
         LBXVE3MS & epi-25OHD3 (nmol/L) & 0 & 0 & 0 & 0 & 0\\
         \hline
    \end{tabular}
    \caption{Application on NHANES 2017-2018 (2): `1' indicates testable and `0' indicates not testable by the Optimal Sample Split with Multiple Tests. For comparison, `*' indicates testable by Bonferroni Correction.}
    \label{tab:application_2}
\end{table}

\section{Conclusion}
\label{s:conclusion}
In this paper, we proposed a novel sensitivity analysis method for treatment effect tests on multiple outcome spaces. 
By leveraging the advantages of data split and multiple hypothesis tests, our approach mitigates sensitivity to treatment assignment bias while extending the applicability of sample split methods to high-dimensional outcome spaces. 
We provided a numerical solution to determine the optimal sample split fraction, ensuring improved test performance.

Through simulations, we demonstrated that \textit{Optimal Sample Split with Multiple Tests}, particularly \textit{Optimal Sample Split FDR Approach}, achieves greater accuracy in high-dimensional outcome spaces with high bias in treatment assignment. 
Additionally, our method performs comparably to existing approaches in low-dimensional settings under minimal bias.

Finally, we applied our proposed method to study the effects of second-hand smoke on children's health using NHANES 2017-2018 data. 
Our results quantitatively identify a broader range of potential health risks associated with second-hand smoke exposure than previous literature had.

\bibliographystyle{plainnat}
\bibliography{reference}

\newpage
\section*{Appendix}
\appendix

\section{Examples of sensitivity model equivalency}
\label{app:sen-equival}
In the main part of the paper, the sensitivity model in treatment assignments bounds the odds ratio of units receiving treatment in any two independent pairs across the entire population by the bias factor $\Gamma \geq 1$ and its reciprocal in Model (\ref{sensitivity_model_odd_ratio}). 

\begin{align}
    \frac{1}{\Gamma} \leq \frac{\pi_{i}(1 - \pi_{j})}{\pi_{j}(1 - \pi_{i})} \leq \Gamma \text{, \, } i = 1, \cdots, I
    \label{sensitivity_model_odd_ratio}
\end{align}
with the corresponding covariates assumed to be $\boldsymbol{x}_i = \boldsymbol{x}_j$. 

The sensitivity model in unobserved covariates is expressed in Model (\ref{sensitivity_model_prob}) with the given parameter $\gamma$. 

\begin{align}
\begin{split}
    \mathrm{Pr}(\mathbf{Z} = \mathbf{z} | \mathcal{F}, \mathcal{Z}) &= \frac{\mathrm{exp}(\gamma \mathbf{z}^T \mathbf{u})}{\sum_{\mathbf{b} \in \mathcal{Z}} \mathrm{exp}(\gamma \mathbf{b}^T \mathbf{u})}\\
    &= \prod_{i = 1}^I \frac{\mathrm{exp}(\gamma \sum_{j = 1}^2 z_{ij} u_{ij})}{\mathrm{exp}(\gamma u_{i1}) + \mathrm{exp}(\gamma u_{i2})}
    \text{\, \, \, for \,} \mathbf{z} \in \mathcal{Z} \text{\, and \,} \mathbf{u} \in \mathcal{U}
    \label{sensitivity_model_prob}
\end{split}
\end{align}
where the unobserved $\mathbf{u} = \left( u_{11}, \cdots, u_{I2} \right)^T$ with $0 \leq u_{ij} \leq 1$, $\mathcal{U} = [0, 1]^{2I} = \{ \mathbf{u} = ( u_{11}, u_{12}, \cdots, u_{I2}) \, | \, i = 1, \cdots, I, j = 1, 2)\}$ and $\gamma = \log(\Gamma)$. 



\begin{itemize}
    \item When $\Gamma = 1$, i.e. $\gamma = \log(\Gamma) = 0$, the sensitivity model aligns with randomized experiments, where there are no unobserved covariates. The treatment assignment distribution $\mathrm{Pr}(\mathbf{Z} = \mathbf{z} | \mathcal{F}, \mathcal{Z})$ is determined to be $\frac{1}{2^I}$. Consequently, the probability of Fisher's sharp null hypothesis $H_0^k$, expressed in the test statistic $T_k$, being rejected $\mathrm{Pr} \left(T_k \geq c | \mathcal{F}, \mathcal{Z} \right)$.
    \item When $\Gamma > 1$, i.e. $\gamma = \log(\Gamma) > 0$, the sensitivity model accommodates scenarios where the uncertainty in treatment assignment distribution $\mathrm{Pr}(\mathbf{Z} = \mathbf{z} | \mathcal{F}, \mathcal{Z})$ is quantified by the bias factor $\Gamma$. The upper and lower bounds of the distribution of treatment assignment $\mathrm{Pr}(\mathbf{Z} = \mathbf{z} | \mathcal{F}, \mathcal{Z})$ for a fixed $\Gamma$ can be computed by optimizing (\ref{sensitivity_model_prob}) over $\mathbf{u} \in \mathcal{U}$, thereby establishing bounds for $\mathrm{Pr} \left(T_k \geq c | \mathcal{F}, \mathcal{Z} \right)$ for the given $\Gamma$.
\end{itemize}

\section{Further explanation of multiple hypothesis tests in the optimal sample split methods}
\label{app:mul-hypo}
To alleviate the overly conservative results given by simple outcome selection, we incorporate multiple hypothesis tests for treatment effect tests on high-dimensional outcome space. 
Instead of only considering the p-value for the most prominent variable, multiple hypothesis tests methods considering family-wise error can effectively control the total type I error across the entire hypothesis tests set, where the family-wise error rate (FWER) is defined as
\begin{equation*}
    \mathrm{FWER} = \mathrm{P}\{\text{any type I error}\} = \mathrm{P}\{\text{any true null hypothesis rejected}\}.
\end{equation*} 
One such method is Bonferroni correction, as introduced by \cite{dunn1961multiple}, ensuring that the family-wise error rate (FWER) remains below a specified level since
\begin{align*}
    \mathrm{FWER} &= \mathrm{P}\{\text{any true null hypothesis rejected}\} \\ 
    &\leq \sum_{k = 1}^K \mathrm{P}\{\text{hypothesis k false rejected}\} \leq \sum_{k = 1}^K \frac{\alpha}{K} = \alpha.
\end{align*}
Similarly, alternative methods within the Bonferroni family, such as those proposed by \cite{holm1979simple} and \cite{hochberg1988sharper}, offer viable options for controlling family-wise error in sample split methods for treatment effect tests. 

However, family-wise error-controlling methods are still contingent on the number of hypotheses, which may significantly influence the results of the pre-selection step. 
Particularly, when dealing with a large number of outcomes (denoted as $K$), Bonferroni correction can become excessively conservative, even for the top promising hypotheses.
Thus, instead of pre-selecting hypotheses before final tests, we advocate for ordering all the hypotheses by their significance, which highlights the prominent hypotheses without being influenced by the number of hypotheses. 
\cite{benjamini1995controlling} proposed the BH procedure, an ordering method in multiple testing that controls the false discovery rate (FDR). 
The FDR is defined as the expected ratio of false positives to the total number of rejections.

\begin{equation*}
    \mathrm{FDR} = \mathrm{E} \left[ \frac{FP}{FP + TP} \right].
\end{equation*}

Ordering methods that control FDR permit a more nuanced approach, allowing Type I errors to be larger than one standard unit, as long as the ratio of total false rejections to total rejections remains relatively small. 
An expanded list of ordering methods that may potentially work for treatment effect tests by sample split can be given as: \cite{benjamini2000adaptive}, \cite{storey2002direct}, \cite{genovese2004stochastic}, \cite{rosenbaum2020conditional}. 

\section{Asymptotic properties of Wilcoxon's signed rank statistic details}
We discuss the details of the asymptotic properties of Wilcoxon’s signed rank statistic in this subsection to provide a more comprehensive understanding of Wilcoxon’s signed rank in treatment effect testing at a large sample size. 
\begin{enumerate}
    \item \textbf{Asymptotic behavior of the critical value:} \\
    Let $T_k$ be the Wilcoxon signed-rank test statistic, denoted as $W_k = w(\mathbf{Z}, \mathbf{R}_k)$. Given a bias factor $\Gamma > 1$ and a large sample size $I$, the critical value $c_\Gamma$ at significance level $\alpha$ is determined by solving:
    \begin{align*} 
        \max_{\mathbf{u} \in \mathcal{U}} \mathrm{Pr}(W_k \geq c_\Gamma \mid \mathcal{F}) = \alpha.
    \end{align*}
    As $I \to \infty$, the critical value $c_\Gamma$ converges to:
    \begin{align*} 
        c_\Gamma \to \frac{\kappa I (I+1)}{2} + \Phi^{-1}(1-\alpha) \sqrt{\frac{\kappa(1-\kappa) I(I+1)(2I+1)}{6}},
    \end{align*}
    where $\kappa = \frac{\Gamma}{1+\Gamma}$ and $\Phi^{-1}(\cdot)$ is the inverse of the standard normal cumulative distribution. For further details, see \cite{rosenbaum1985bias} and \cite{heller2009split}.
    
    \item \textbf{Asymptotic behavior of test power in randomized experiments:} \\
    Define the treated-minus-control difference for matched pair $i$ and outcome $k$ as:
    \begin{align*}
        D_{ik} = (2 Z_{i1} - 1)(R_{i1k} - R_{i2k}).
    \end{align*}
    The probabilities associated with $D_{ik}$ are:
    \begin{itemize}
        \item $p_{0k} = \mathrm{Pr}(D_{ik} > 0)$,
        \item $p_{1k} = \mathrm{Pr}(D_{ik} + D_{jk} > 0)$,
        \item $p_{2k} = \mathrm{Pr}(D_{ik} + D_{jk} > 0, D_{ik} + D_{sk} > 0)$ for $0 \leq i < j < s \leq I$.
    \end{itemize}
    Assuming independent and identically distributed (i.i.d.) $D_{ik}$ and no unmeasured confounding, as $I \to \infty$, the power of Wilcoxon's test converges to:
    \begin{align*}
        \mathrm{Pr}(W_k \geq c_\Gamma \mid \mathcal{F}) \to 1 - \Phi\left(\frac{c_\Gamma - \mu_k}{\sigma_k}\right),
    \end{align*}
    where the mean and variance of $W_k$ are given by:
    \begin{align*}
        \mu_k &= \frac{1}{2} I (I-1)p_{1k} + Ip_{0k},\\
        \sigma^2_k &= I (I-1)(I-2)(p_{2k} - p_{1k}^2) \\
        &+ \frac{1}{2} I (I-1)p_{1k} \left[ 2(p_{0k} - p_{1k})^2 + 3p_{1k}(1 - p_{1k})\right] + Ip_{0k} (1 - p_{0k}).
    \end{align*}
    For details, see \cite{10.2307/27643648}, \cite{hettmansperger1978statistical}, and \cite{shieh2007power}.
    
    \item \textbf{Asymptotic behavior of test power with unmeasured confounders:} \\
    Relaxing the assumption of no unmeasured confounding, we analyze the sensitivity model with bias factor $\Gamma$. The quantile $\frac{c_{\Gamma} - \mu_k}{\sigma_k}$ can be further approximated as:
    \begin{align*}
        \frac{c_{\Gamma} - \mu_k}{\sigma_k} &= \left( \frac{\kappa I (I + 1)}{2} + \Phi^{-1} (1 - \alpha) \sqrt{\frac{\kappa (1 - \kappa) I (I + 1) (2I + 1)}{6}} - \frac{I (I - 1) p_{1k}}{2} - I p_{0k} \right)\\
        & \times \biggl( I (I - 1) (I - 2) (p_{2k} - p_{1k}^2) + \frac{I (I - 1)}{2} \left[ 2 (p_{0k} - p_{1k})^2 + 3 p_{1k} (1 - p_{1k}) \right] \\
        &+ I p_{0k} (1 - p_{0k}) \biggl)^{-\frac{1}{2}}.
    \end{align*}
    As $I \to \infty$, the test power under unmeasured confounding converges to:
    \begin{align*}
        \mathrm{Pr}(W_k \geq c_\Gamma \mid \mathcal{F}) \to 
        \begin{cases}
            1, & \text{if } \Gamma < \widetilde{\Gamma}, \\
            0, & \text{if } \Gamma > \widetilde{\Gamma},
        \end{cases}
    \end{align*}
    Here, the threshold $\widetilde{\Gamma}$ is defined as \textbf{design sensitivity}. Under the setting of Wilcoxon’s Signed Rank Statistic, it is computed to be $\frac{p_{1k}}{1 - p_{1k}}$. 
    The design sensitivity $\widetilde{\Gamma}$ determines the robustness of the test against unmeasured confounders. If $\Gamma < \widetilde{\Gamma}$, the test retains power asymptotically; otherwise, power vanishes. See \cite{rosenbaum2004design} for further discussion.
\end{enumerate}

\section{Simulation Results} 
\label{sim:tables}
Similarly to what we have presented in Section 4 of the main part of the paper, the simulation results for the case that the outcome numbers $K = 10, 100, 500$ are provided in the Tables \ref{tab:sim-k-10}, \ref{tab:sim-k-100}, and \ref{tab:sim-k-500}, respectively. The explanation and analysis of these results can also be found in the Method Section of the main part of the paper.

\begin{table}[]
\centering
\begin{tabular}{llllll}
\hline
      & & \multicolumn{4}{c}{I}\\
      & & 100    & 200   & 500   & 1000\\ 
      \cline{3-6} 
      Gamma & & & & &\\ 
      \hline
      1 & Bonferroni                & 0.1 & 0.34 & 0.92 & 1\\
      & Naive split                 & 0.24 &          0.4 &         0.87 &            1 \\
      &                             & [0.53, 0.99] &   [0.4, 0.8] & [0.48, 0.72] & [0.39, 0.73] \\
      & Optimal split selection  & 0.24 &         0.39 &         0.86 &            1 \\
      &                             & [0.99, 0.99] &   [0.4, 0.8] & [0.48, 0.78] &  [0.4, 0.74] \\
      & Optimal split rank   & 0.28 &         0.47 &         0.95 &            1 \\
      &                             & [0.95, 0.96] & [0.95, 0.99] & [0.91, 0.99] & [0.63, 0.99] \\
      \hline
      1.25 & Bonferroni             & 0.68 & 0.98 & 1 & 1\\
      & Naive split                 & 0.76 &         0.99 &            1 &            1 \\
      &                             & [0.55, 0.83] & [0.44, 0.76] &  [0.2, 0.92] & [0.12, 0.96] \\
      & Optimal split selection  & 0.76 &         0.99 &            1 &            1 \\
      &                             & [0.55, 0.83] & [0.44, 0.78] &  [0.2, 0.92] & [0.12, 0.96] \\
      & Optimal split rank   & 0.75 &         0.99 &            1 &            1 \\
      &                             & [0.9, 0.98] & [0.77, 0.99] & [0.31, 0.99] & [0.17, 0.99] \\
      \hline
      1.5 & Bonferroni              & 0 & 0 & 0 & 0\\
      & Naive split                 & 0.02 &         0.04 &         0.02 &         0.03 \\
      &                             & [0.05, 0.41] & [0.05, 0.05] & [0.01, 0.19] & [0.05, 0.05] \\
      & Optimal split selection  & 0.02 &         0.04 &         0.02 &         0.03 \\
      &                             & [0.05, 0.41] & [0.05, 0.05] & [0.01, 0.19] & [0.05, 0.05] \\
      & Optimal split rank   & 0.03 &         0.01 &         0.01 &         0.01 \\
      &                             & [0.15, 0.15] & [0.05, 0.81] & [0.38, 0.38] & [0.04, 0.04] \\
      \hline
      1.75 & Bonferroni             & 0 & 0 & 0 & 0\\
      & Naive split                 & 0.01 &         0.02 &         0.02 &         0.01 \\
      &                             & [0.06, 0.46] & [0.07, 0.09] & [0.02, 0.02] & [0.02, 0.02] \\
      & Optimal split selection  & 0.01 &         0.02 &         0.02 &         0.01 \\
      &                             & [0.06, 0.46] & [0.07, 0.09] & [0.02, 0.02] & [0.02, 0.02] \\
      & Optimal split rank   & 0.01 &         0.01 &            0 &            0 \\
      &                             & [0.09, 0.31] & [0.15, 0.36] & [0.01, 0.99] & [0.01, 0.99] \\
      \hline
      2 & Bonferroni                & 0.24 & 0.55 & 0.99 & 1\\
      & Naive split                 & 0.57 &         0.84 &            1 &            1 \\
      &                             & [0.67, 0.86] & [0.76, 0.86] & [0.49, 0.92] & [0.23, 0.97] \\
      & Optimal split selection  & 0.57 &         0.84 &            1 &            1 \\
      &                             & [0.67, 0.86] & [0.76, 0.86] & [0.49, 0.92] & [0.23, 0.97] \\
      & Optimal split rank   & 0.33 &         0.62 &            1 &            1 \\
      &                             & [0.96, 0.97] & [0.95, 0.99] & [0.75, 0.99] & [0.39, 0.99] \\
      \hline
\end{tabular}
\caption{Simulation result, when the number of outcomes $K = 10$: The table presents the test powers and the corresponding recommended sample split fraction, set at the given number of outcomes $K = 10$ for each method under various bias factors and sample sizes. The simulation is replicated $1000$ times with the significance level to be $\alpha = 0.05$. The first row for each simulation setting presents the test power, while the brackets in the second row (for the split methods) indicate the recommended set for the split fraction $\zeta$.}
\label{tab:sim-k-10}
\end{table}

\begin{table}[]
\centering
\begin{tabular}{llllll}
\hline
      & & \multicolumn{4}{c}{I}\\
      & & 100    & 200   & 500   & 1000\\ 
      \cline{3-6} 
      Gamma & & & & &\\ 
      \hline
      1 & Bonferroni                & 0.147 & 0.269 & 0.382 & 0.442\\
      & Naive split                 & 0.081 &        0.098 &          0.1 &          0.1 \\
      &                             & [0.45, 0.73] & [0.44, 0.74] & [0.14, 0.86] & [0.11, 0.94] \\
      & Optimal split selection  & 0.135 &        0.245 &        0.357 &        0.419 \\
      &                             & [0.54, 0.65] & [0.56, 0.67] & [0.46, 0.75] & [0.42, 0.76] \\
      & Optimal split rank   & 0.28 &        0.379 &        0.516 &        0.599 \\
      &                             & [0.93, 0.97] & [0.88, 0.99] & [0.92, 0.99] & [0.93, 0.99] \\
      \hline
      1.25 & Bonferroni             & 0.175 & 0.276 & 0.474 & 0.650\\
      & Naive split                 & 0.096 &          0.1 &          0.1 &          0.1 \\
      &                             & [0.43, 0.7] & [0.27, 0.87] & [0.09, 0.96] & [0.04, 0.97] \\
      & Optimal split selection  & 0.167 &        0.268 &        0.432 &        0.612 \\
      &                             & [0.5, 0.71] &  [0.5, 0.66] &   [0.5, 0.7] & [0.48, 0.72] \\
      & Optimal split rank   & 0.298 &        0.438 &        0.618 &         0.69 \\
      &                             & [0.92, 0.97] & [0.97, 0.98] & [0.97, 0.99] & [0.79, 0.99] \\
      \hline
      1.5 & Bonferroni              & 0.045 & 0.164 & 0.359 & 0.496\\
      & Naive split                 & 0.066 &        0.092 &          0.1 &          0.1 \\
      &                             & [0.64, 0.73] & [0.67, 0.83] & [0.39, 0.93] & [0.15, 0.97] \\
      & Optimal split selection  & 0.078 &        0.168 &        0.324 &        0.472 \\
      &                             & [0.64, 0.64] & [0.55, 0.71] & [0.52, 0.73] & [0.46, 0.72] \\
      & Optimal split rank   & 0.161 &        0.317 &        0.474 &         0.57 \\
      &                             & [0.96, 0.97] & [0.97, 0.99] & [0.99, 0.99] & [0.94, 0.99] \\
      \hline
      1.75 & Bonferroni             & 0.001 & 0.007 & 0.029 & 0.091\\
      & Naive split                 & 0.021 &        0.035 &        0.082 &        0.099 \\
      &                             & [0.43, 0.68] & [0.54, 0.75] &   [0.6, 0.8] & [0.45, 0.81] \\
      & Optimal split selection  & 0.02 &        0.037 &        0.083 &        0.104 \\
      &                             & [0.43, 0.68] & [0.54, 0.75] &  [0.79, 0.8] & [0.49, 0.81] \\
      & Optimal split rank   & 0.022 &        0.033 &        0.074 &        0.125 \\
      &                             & [0.91, 0.91] & [0.99, 0.99] & [0.99, 0.99] & [0.99, 0.99] \\
      \hline
      2 & Bonferroni                & 0.097 & 0.111 & 0.184 & 0.201\\
      & Naive split                 & 0.098 &          0.1 &          0.1 &          0.1 \\
      &                             & [0.43, 0.71] & [0.18, 0.82] & [0.08, 0.94] & [0.05, 0.97] \\
      & Optimal split selection  & 0.105 &        0.123 &        0.189 &        0.203 \\
      &                             & [0.45, 0.77] & [0.31, 0.78] & [0.47, 0.63] & [0.32, 0.79] \\
      & Optimal split rank   & 0.12 &        0.151 &        0.202 &        0.218 \\
      &                             & [0.96, 0.97] & [0.96, 0.99] & [0.92, 0.99] & [0.94, 0.99] \\
      \hline
\end{tabular}
\caption{Simulation result, when the numbers of outcomes $K = 100$: The table presents the test powers and the corresponding recommended sample split fraction, sets at the given number of outcomes $K = 100$ for each method under various bias factors and sample sizes. The simulation is replicated $1000$ times with the significant level to be $\alpha = 0.05$. The first row for each simulation setting presents the test power, while the brackets in the second row (for the split methods) indicate the recommended set for the split fraction $\zeta$}.
\label{tab:sim-k-100}
\end{table}

\begin{table}[]
\centering
\begin{tabular}{llllll}
\hline
      & & \multicolumn{4}{c}{I}\\
      & & 100    & 200   & 500   & 1000\\ 
      \cline{3-6} 
      Gamma & & & & &\\ 
      \hline
      1 & Bonferroni                & 0.226 & 0.371 & 0.585 & 0.694\\
      & Naive split                 & 0.02 &         0.02 &         0.02 &         0.02 \\
      &                             & [0.33, 0.83] & [0.16, 0.92] & [0.07, 0.97] & [0.03, 0.98] \\
      & Optimal split selection  & 0.1922 &       0.3316 &         0.55 &       0.6564 \\
      &                             & [0.6, 0.73] & [0.59, 0.73] & [0.54, 0.77] & [0.46, 0.81] \\
      & Optimal split rank   & 0.4712 &       0.5694 &       0.7412 &       0.8172 \\
      &                             & [0.91, 0.97] & [0.89, 0.99] & [0.87, 0.99] & [0.83, 0.99] \\
      \hline
      1.25 & Bonferroni             & 0.080 & 0.185 & 0.355 & 0.434\\
      & Naive split                 & 0.0198 &         0.02 &         0.02 &         0.02 \\
      &                             & [0.35, 0.81] & [0.16, 0.92] & [0.05, 0.97] & [0.03, 0.98] \\
      & Optimal split selection  & 0.0884 &       0.1744 &       0.3356 &       0.4248 \\
      &                             & [0.58, 0.74] & [0.58, 0.75] & [0.55, 0.74] & [0.44, 0.79] \\
      & Optimal split rank   & 0.2274 &       0.3314 &       0.4378 &        0.476 \\
      &                             & [0.95, 0.98] & [0.95, 0.99] & [0.93, 0.99] & [0.79, 0.99] \\
      \hline
      1.5 & Bonferroni              & 0.064 & 0.155 & 0.255 & 0.342\\
      & Naive split                 & 0.0196 &         0.02 &         0.02 &         0.02 \\
      &                             & [0.5, 0.76] &  [0.21, 0.9] & [0.08, 0.96] & [0.04, 0.98] \\
      & Optimal split selection  & 0.0762 &       0.1554 &       0.2496 &       0.3312 \\
      &                             & [0.62, 0.69] & [0.56, 0.71] & [0.53, 0.76] & [0.46, 0.76] \\
      & Optimal split rank   & 0.16 &       0.2674 &       0.3476 &       0.3892 \\
      &                             & [0.95, 0.99] & [0.98, 0.99] & [0.96, 0.99] & [0.93, 0.99] \\
      \hline
      1.75 & Bonferroni             & 0.003 & 0.030 & 0.129 & 0.218\\
      & Naive split                 & 0.0128 &       0.0182 &         0.02 &         0.02 \\
      &                             & [0.77, 0.77] & [0.53, 0.85] &  [0.2, 0.92] &  [0.1, 0.97] \\
      & Optimal split selection  & 0.016 &       0.0434 &       0.1346 &        0.211 \\
      &                             & [0.63, 0.77] & [0.56, 0.72] & [0.61, 0.68] & [0.51, 0.73] \\
      & Optimal split rank   & 0.0512 &       0.1098 &       0.2132 &       0.2602 \\
      &                             & [0.96, 0.96] & [0.98, 0.99] & [0.99, 0.99] & [0.97, 0.99] \\
      \hline
      2 & Bonferroni                & 0.016 & 0.057 & 0.130 & 0.156\\
      & Naive split                 & 0.0186 &         0.02 &         0.02 &         0.02 \\
      &                             & [0.59, 0.7] &  [0.27, 0.9] & [0.09, 0.97] & [0.06, 0.98] \\
      & Optimal split selection  & 0.0316 &       0.0694 &       0.1326 &        0.157 \\
      &                             & [0.57, 0.69] & [0.54, 0.72] & [0.47, 0.72] & [0.43, 0.79] \\
      & Optimal split rank   & 0.0576 &       0.1056 &       0.1572 &       0.1678 \\
      &                             & [0.97, 0.98] & [0.98, 0.99] & [0.98, 0.99] & [0.75, 0.99] \\
      \hline
\end{tabular}
\caption{Simulation result, when the number of outcomes $K = 500$: The table presents the test powers and the corresponding recommended sample split fraction, set at the given number of outcomes $K = 500$ for each method under various bias factors and sample sizes. The simulation is replicated $1000$ times with the significance level to be $\alpha = 0.05$. The first row for each simulation setting presents the test power, while the brackets in the second row (for the split methods) indicate the recommended set for the split fraction $\zeta$}.
\label{tab:sim-k-500}
\end{table}

\section{Application Dataset Matched Pairs}
\begin{figure}
    \centering
    \includegraphics[width=0.8\linewidth]{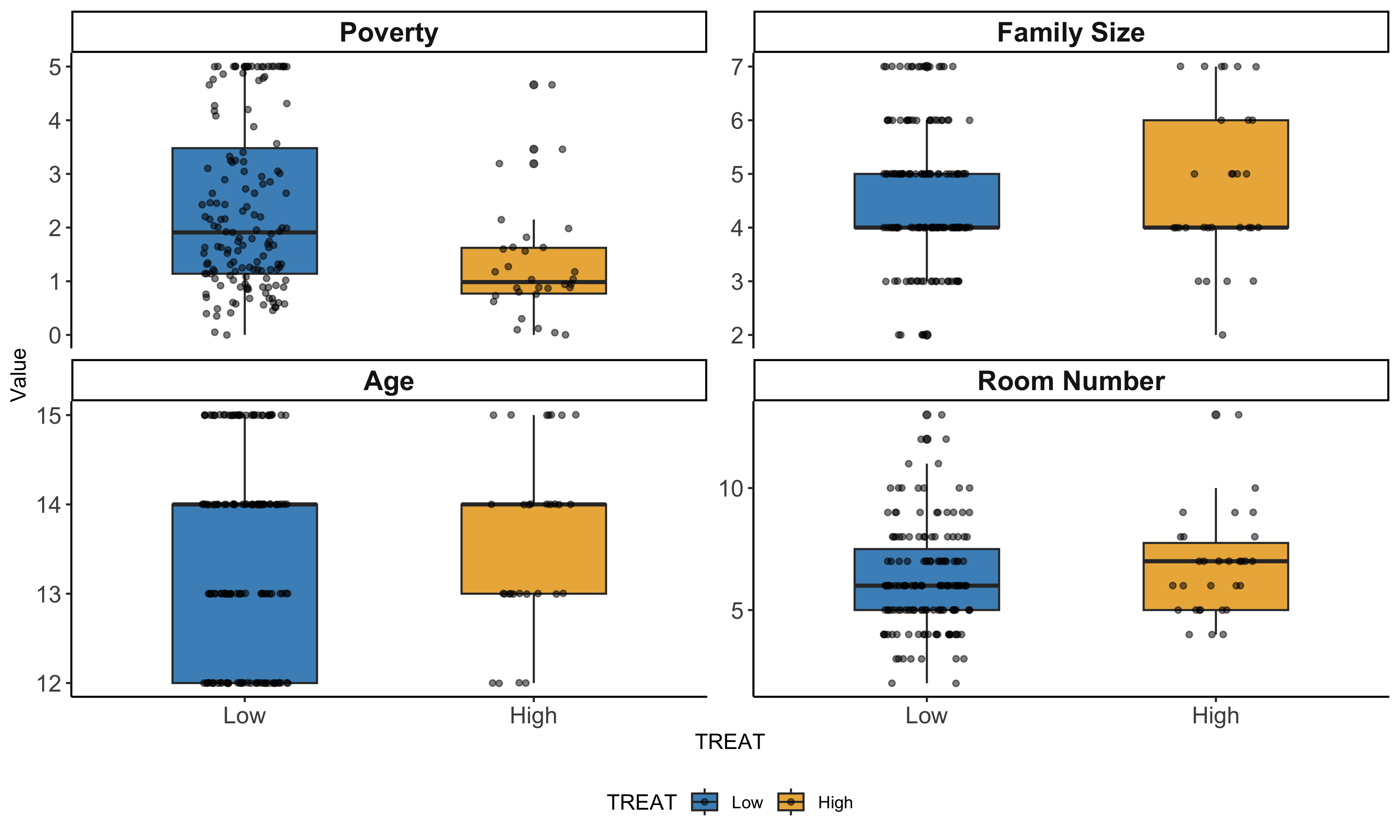}
    \caption{Box plot for the matched nonbinary covariates - age, poverty, number of rooms in the house, and size of the family - between treated and controlled groups.}
    \label{fig:pretreatment_covariates}
\end{figure}

Following the Application Section of the main paper, we provide the distribution of the matched pairs after performing full matching based on education, gender, race/ethnicity, age, poverty level, number of rooms in the house, and family size on the dataset from the US National Health and Nutrition Examination Survey (NHANES) 2017-2018 at \cite{nhanes2017overview} in Figure \ref{fig:pretreatment_covariates}.

The balance of distribution among covariates of the treated and control groups shows the effectiveness of the full matching method and the comparability of the treated-minus-control difference under the known covariates.

\end{document}